# Legibility of Videos with ASL signers

Raja S. Kushalnagar

Gallaudet University
800 Florida Ave NE, Washington, DC 20002 USA
raja.kushalnagar@gallaudet.edu

## Abstract

The viewing size of a signer correlates with legibility, i.e., the ease with which a viewer can recognize individual signs. The WCAG 2.0 guidelines (G54) mention in the notes that there should be a mechanism to adjust the size to ensure the signer is discernible but does not state minimum discernibility guidelines. The fluent range (the range over which sign viewers can follow the signers at maximum speed) extends from about 7° to 20°, which is far greater than 2° for print. Assuming a standard viewing distance of 16 inches from a 5-inch smartphone display, the corresponding sizes are from 2 to 5 inches, i.e., from 1/3$^{rd}$ to full-screen. This is consistent with vision science findings about human visual processing properties, and how they play a dominant role in constraining the distribution of signer sizes.

## Introduction

Signing is of fundamental importance in modern culture and communication among signers, primarily those who are deaf and hard of hearing. There are over 500,000 people who rely on sign language in the United States (Mitchell, Young, Bachleda, & Karchmer, 2006) and over 70 million people worldwide (WFD, 2017). Many deaf people prefer to communicate via sign language over English, as sign language is fully visual compared with English. This has led to the development of Deaf culture, a set of social beliefs, behaviors, art, literary traditions, history, values, and shared institutions of communities that are influenced by deafness and which use sign languages as the main means of communication. The advent of signed videos is comparable to the impact of the printing press in facilitating cultural transmission in cultures world-wide. Deaf people associate in closed, close-knit local communities linked to schools or locations. They are united by their common use of sign language, which functions as a symbol of identity, medium of interaction and basis of cultural knowledge (Baker-Shenk & Cokely, 1996; Hauser, O'Hearn, McKee, Steider, & Thew, 2010).

ASL communication is conveyed not only by hand movements, but also from head movements (e.g. head tilts), facial expressions and use of space around the body (Bellugi & Fischer, 1972). Signs are produced in a very specific body region defined from the waist, with reach of slightly bent elbows to the top of head, and from the location of the sign in space. The viewing size of the signer is a crucial factor determining the signer's legibility, i.e., the ease with which a viewer can recognize individual signs, not including finger-spelling. Sign legibility differs from sign readability, which is the ease with which a viewer can recognize signed narration.

When sign communication is conveyed over video, many characteristics of the sign communication between signer and viewer change due to technical constraints. Signing videos create a new relationship between signers and viewers in content delivery (Keating & Mirus, 2003, 2004), and that signers using cameras alter the spatial dimension of sign language, from three dimensions in face-to-face communication to two dimensions in video. Signers changed their body positions to accommodate the narrow signing window, due to the camera field-of-view constraints, compared with the space available in face-to-face communication. Signers changed to optimize the display of their bodies, face, and hands.

Other factors that impact the ease of understanding a signer include attributes such as the viewing angle between viewer and signer, the signer's signing space, the color of the signer's clothes, etc. Research has shown that sign language interpreter comprehension on television appears to be problematic, in part due to the small size of the interpreter picture and fast signing (Ofcom, 2007; Xiao & Li, 2013).

## Size

Sign legibility has implicitly been driven by the properties of human vision, as evidenced by findings that writing symbols have evolved in response to visual pressures rather than writing pressures. Studies (Changizi, Zhang, Ye, & Shimojo, 2006; Morin, 2018) have found that legibility is related to contour configuration and size. Specifically, the distribution of contour topological configurations and sizes that are common in writing systems corresponds to the distribution found in natural images, implying that written forms have been designed to take advantage of visual mechanisms that evolved for perception in the natural world.

Unlike speech which is characterized by rapid and sequential order of words, sign language is characterized by slower production and simultaneous layering of components spread over the signing space of the body (Wilbur, 1999; Wilbur & Allen, 1991). Furthermore, the layering effect implies that the articulation of each piece of information cannot interfere with the perception and production of the others, and that the signing space has to be more clearly visible. In other words, signs have evolved to ensure that their units are clearly distinguishable.

Visual acuity (ability to distinguish visual differences) is sharpest within about 10 degrees range around the center of focus (Mandelbaum & Sloan, 1947). The angular size of visual information, including signers, is measured in minutes of arc or degrees of visual angle from the center of focus. The angular measure depends on both the physical size of the signer and the subject's viewing distance, which gives a more accurate measure of human perception of perceived visual size. The angular size is correlated with the viewing size in the eye, which is known as the retinal image size and is defined as follows:

Angular size (in degrees) = 57.3 × physical size / viewing distance

The physical size and viewing distance are measured in the same units, such as millimeters, centimeters, or inches. This equation is an approximation, which holds when the physical size is significantly smaller than the viewing distance.

For instance, suppose the height of a displayed signer in a phone display is 14 mm (physical size), and the viewer keeps the phone at a typical viewing distance of 40 cm (16 inches). The angular height of the signer at the eye is 2°. If the reader reduces the viewing distance from 40 cm to 20 cm, the angular character size doubles to 4°, but the physical character size, of course, does not change. In clinical vision applications, angular character size is often expressed using metrics from visual acuity testing including Snellen notation (Sloan, 1951) or logMAR (Rosser, 2001).

# Methodology

We collected and analyzed 240 videos that had one or more signers narrating information via American Sign Language. Of the 240 videos, half (120) were selected from DeafVideo.TV and the other half were from YouTube. YouTube is a general video-streaming service, while DeafVideo.TV is a video-streaming service specifically aimed at deaf signers and the Deaf community.

We selected videos in which the contributor was a fluent deaf signer adult that used ASL and had uploaded at least 50 posts to either video-streaming service. For all videos, we gathered video resolution and signing size, and assumed a standard viewing distance of 16 inches from a 5.5-inch phone display (Bababekova, Rosenfield, Hue, & Huang, 2011).

## Data collection

The videos were assessed for quantitative and qualitative factors, addressing the research questions. The following quantitative factors were measured:
1. *Signing size.* The signing size was determined by measuring the size of the signer in relation to the size of the entire video.
2. *Signing rate.* Signing rate was determined by counting signs in of the video and then normalizing the count per second.
3. *Average length* (and standard deviation) of posts in seconds.

## Filtering

To determine whether the contributor was a fluent signer, we looked for a statement from the video creator identifying themselves as a fluent deaf signer. This was most often found in the Deafvideo.TV profile or one of the YouTube videos. If text was included in the original video, it was excluded from the study to ensure that all content was ASL only.

On DeafVideo.TV, all videos were in American Sign Language, and the history of all of postings could be easily tracked on this site. Identifying ASL videos on YouTube was more complex as YouTube contains a very large range of audio-visual content with only some ASL content. ASL videos were found on YouTube by using the keyword, "ASL" as there was no American Sign Language or Deaf community on YouTube.

# Results

The DeafVideo.TV distributions were clustered with a mean visual angle of 16° in a range from 7° to 16° (1.92 to 4.4 inches on a 5.5-inch tall display). These ranges indicate that the video creators judged the videos to be legible at around one third the screen size. For YouTube videos, the mean visual angle was 18°, and distribution is clustered like DeafVideo.TV but with a somewhat smaller range of 9-19° (2.47– 5.22 inches on a 5.5-inch tall display). These results indicate that the video creators judged the videos to be legible at around half the screen size.

To gauge whether there was a correlation between signer size and video length, a repeated measures ANOVA was carried out for the average length of videos. There was no significant difference between the videos or between YouTube and DeafVideo.TV. Even though there is no length limit in YouTube videos, there were no instances of videos over 6 minutes long (360 seconds).

# Conclusions

One of the criterions required for meeting the WCAG 2.1 AAA standard is to provide sign language interpretation. Due to lack of access to spoken communication in a predominantly spoken world, a significant fraction of deaf individuals are not proficient in their communities' spoken language and read at a 4[th] grade level or less (Allen, 1986; Traxler, 2000) and may be more proficient in sign language compared with their communities' spoken language. Furthermore, some deaf people prefer to communicate via sign language over their communities' spoken language, as sign language is visually accessible compared with English, and intonation, emotion and other audio information is shown in sign language interpretation, but not in captions. As a result, sign language interpretation provides richer and more equivalent visual access to the audio component of synchronized media.

When sign communication is conveyed over video, many characteristics of the signed information conveyance change due to technical constraints, primarily due to the change from three-dimensions to two-dimensions. WGAG 2.1, success criterion 1.2.6 (Sign Language, pre-recorded) states that sign language interpretation is provided for all pre-recorded audio content in synchronized media. The associated guideline, G54, addresses issues related to the synchronization, such as an inlay of a second video stream of the signer in the main video. It notes that if the video stream is too small, the sign language interpreter will be indiscernible. When creating a video steam that includes a video of a sign language interpreter, creators are asked to make sure there is a mechanism to play the video stream full screen in the accessibility-supported content technology. Otherwise, they are asked to be sure the interpreter portion of the video is adjustable to the size it would be had the entire video stream been full screen.

The G.54 guideline does not specify the minimum size of the video with the signer, relative to the average viewing distance. The findings from our analysis of 240 videos by native signers on both YouTube and DeafVideo.TV suggests that at normal viewing distance, the minimum viewing size of a signer should be at least 1/3[rd] the size of the full screen, and preferably larger.

# References

Allen, T. E. (1986). Patterns of academic achievement among hearing impaired students: 1974 and 1983. *Deaf Children in America*, 161–206.

Bababekova, Y., Rosenfield, M., Hue, J. E., & Huang, R. R. (2011). Font Size and Viewing Distance of Handheld Smart Phones. *Optometry and Vision Science*, *88*(7), 795–797. https://doi.org/10.1097/OPX.0b013e3182198792

Baker-Shenk, C., & Cokely, D. (1996). *American Sign Language, A Teacher's Resource Text on Grammar and Culture*. Gallaudet University Press.

Bellugi, U., & Fischer, S. (1972). A comparison of sign language and spoken language. *Cognition*, *1*(2–3), 173–200. https://doi.org/10.1016/0010-0277(72)90018-2

Changizi, M. A., Zhang, Q., Ye, H., & Shimojo, S. (2006). The Structures of Letters and Symbols throughout Human History Are Selected to Match Those Found in Objects in Natural Scenes. *The American Naturalist*, *167*(5), E117–E139. https://doi.org/10.1086/502806

Hauser, P. C., O'Hearn, A., McKee, M., Steider, A., & Thew, D. (2010). DEAF EPISTEMOLOGY: DEAFHOOD AND DEAFNESS. *American Annals of the Deaf*, *154*(5), 486. Retrieved from http://search.proquest.com/docview/288323505?accountid=108 LA - English

Keating, E., & Mirus, G. (2003). Examining Interactions across Language Modalities: Deaf Children and Hearing Peers at School. *Anthropology <html_ent Glyph="@amp;" Ascii="&"/> Education Quarterly*, *34*(2), 115–135. https://doi.org/10.1525/aeq.2003.34.2.115

Keating, E., & Mirus, G. (2004). American Sign Language in virtual space: Interactions between deaf users of computer-mediated video communication and the impact of technology on language practices. *Language in Society*, *32*(05), 693–714. https://doi.org/10.1017/S0047404503325047

Mandelbaum, J., & Sloan, L. L. (1947). Peripheral visual acuity with special reference to scotopic illumination. *American Journal of Ophthalmology*, *30*(5), 581–588. Retrieved from http://www.ncbi.nlm.nih.gov/pubmed/20241713

Mitchell, R. E., Young, T. A., Bachleda, B., & Karchmer, M. A. (2006). How Many People Use ASL in the United States? Why Estimates Need Updating. *Sign Language Studies*, *6*(3), 306–335. https://doi.org/10.1353/sls.2006.0019

Morin, O. (2018). Spontaneous Emergence of Legibility in Writing Systems: The Case of Orientation Anisotropy. *Cognitive Science*, *42*(2), 664–677. https://doi.org/10.1111/cogs.12550

Ofcom. (2007). *Signing on television: New arrangements for low audience channels*.

Rosser, D. A. (2001). The development of a "reduced logMAR" visual acuity chart for use in routine clinical practice. *British Journal of Ophthalmology*, *85*(4), 432–436. https://doi.org/10.1136/bjo.85.4.432

Sloan, L. L. (1951). MEASUREMENT OF VISUAL ACUITY. *A.M.A. Archives of Ophthalmology*, *45*(6), 704. https://doi.org/10.1001/archopht.1951.01700010719013

Traxler, C. B. (2000). The Stanford Achievement Test, 9th Edition: National Norming and Performance Standards for Deaf and Hard-of-Hearing Students. *Journal of Deaf Studies and Deaf Education*, *5*(4), 337–348. https://doi.org/10.1093/deafed/5.4.337

WFD. (2017). Sign Language. Retrieved January 7, 2018, from https://wfdeaf.org/human-rights/crpd/sign-language

Wilbur, R. B. (1999). Stress in ASL: Empirical Evidence and Linguistic Issues. *Language and Speech*, *42*(2–3), 229–250. https://doi.org/10.1177/00238309990420020501

Wilbur, R. B., & Allen, G. D. (1991). Perceptual Evidence Against Internal Structure in American Sign Language Syllables. *Language and Speech*, *34*(1), 27–46. https://doi.org/10.1177/002383099103400102

Xiao, X., & Li, F. (2013). Sign language interpreting on Chinese TV: a survey on user perspectives. *Perspectives*, *21*(1), 100–116. https://doi.org/10.1080/0907676X.2011.632690